\documentclass{PoS}

\def\mpi2{m_\pi^2}
\def\mK2{m_K^2}

\newcommand{\bea}{\begin{eqnarray}}
\newcommand{\eea}{\end{eqnarray}}
\newcommand{\be}{\begin{equation}}
\newcommand{\ee}{\end{equation}}

\newsavebox{\DERIVBOXZLM}
\savebox{\DERIVBOXZLM}[2.5em]{$\Longrightarrow\hspace{-1.5em}
\raisebox{.2ex}{*}
\hspace{-.7em}\raisebox{-.8ex}{\scriptsize lm}\hspace{.7em}$}

\usepackage{fleqn,epsf}
\newcommand{\newCED}[2]{{#2}}
\title{Contribution of charm anihilation to the hyperfine splitting in charmonium in the
quenched case}

\ShortTitle{Contributions of charm anihilation to the hyperfine splitting in charmonium}

\author{\speaker{L. Levkova} and C. DeTar\\
        University of Utah,  Salt Lake City, UT 84112, USA\\
        E-mail: \email{ludmila@phys.columbia.edu, detar@physics.utah.edu}}
\author{Fermilab Lattice and MILC Collaborations}

\abstract{In calculations of the hyperfine splitting in charmonium, 
the contributions of the disconnected diagrams
is considered small and is typically ignored. We aim to estimate nonperturbatively the size of
the resulting error, 
which could potentially affect the high precision calculations of the charmonium
spectrum. Following our work on the effects of the disconnected diagrams 
in unquenched QCD presented at Lattice 2007, we study the same problem in the 
quenched case. On dynamical ensembles the disconnected charmonium propagators 
contain light modes which complicate the extraction of the signal at large distances.
In the \newCED{}{fully} quenched case, where there are no such light modes, the interpretation
of the signal is simplified. We present results from lattices with $a\approx 0.09$ fm 
and $a\approx 0.063$ fm.}

\FullConference{The XXVI International Symposium on Lattice Field Theory \\
                 July 14 - 19, 2008\\
                 Williamsburg, Virginia, USA}

\begin{document}
\section{Introduction}
Lattice calculations of the hyperfine splitting in charmonium usually 
ignore the contributions of the disconnected diagrams. This  simplification 
leads to an error and our goal is to determine 
its degree and thus elucidate the origins of the 
the current discrepancies between the lattice calculations and the experimental 
value of the hyperfine splitting of $117$ MeV. The discrepancy, which even for improved 
actions \cite{on,hisq} is within $10\%$ below the experimental
value, could be a result both of the omission of the disconnected diagrams  and
of the discretization errors in the heavy-quark actions. 
Our work on dynamical lattices reported in \cite{mine}, improved over previous
attempts \cite{fs1,fs2} to determine the effect of the disconnected diagrams by going to
finer lattice spacings, larger volumes, using unbiased subtraction for the
stochastic estimation of the operator traces, and most importantly, using point-to-point
({\it ptp}) propagators. This allowed us to estimate that the contribution of the 
disconnected
diagrams decreases the lattice hyperfine splitting, which implied
that the main culprit for the discrepancy between lattice and experiment is the discretization error 
in the heavy-quark action. Our estimation was based on fits of the disconnected 
{\it ptp} propagators to an asymptotic formula
which did not take into account the rotational symmetry violations visible in
our data. Here we introduce a new procedure
which should take these effects into account and we study the effect on quenched lattices,
where complications from light-quark intermediate states are absent. 

\begin{figure}[ht]
\begin{center}
\includegraphics[width=0.85\textwidth]{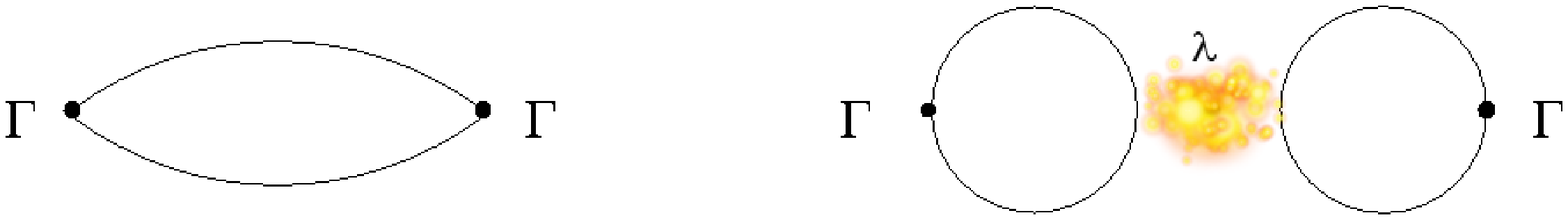}
\caption{Connected (left) and disconnected (right) diagrams contributing 
to the full propagator on lattices quenched with respect to the charm
quark.}
\label{fig:diagram}
\end{center}
\end{figure}

\newCED{The full charmonium propagator, $F(t)$,
is a sum of two contributions: connected, $C(t)$ and disconnected, $D(t)$:
\be
F(t) = C(t) + D(t) = \sum_n\langle 0|O|n \rangle\langle n|O|0 \rangle e^{-E_nt}.
\ee
Diagrammatically they are shown in Fig.~\ref{fig:diagram}, where $D(t)$ is represented by
just one hairpin diagram due to the fact that there are no charm loops in the sea.}
{
The full charmonium propagator, $F(t)$, is a sum of two contributions,
connected, $C(t)$, and disconnected, $D(t)$, shown in Fig.~\ref{fig:diagram}:
\be
F(t) = C(t) + D(t) = \sum_n\langle 0|O|n \rangle\langle n|O|0 \rangle e^{-E_nt}.
\ee
The mass shift due to charm quark loops can be treated as a
perturbation, in which case, to first order, both contributions can be
computed without charmed sea quarks.
} 
The operator $O$ is defined to be
hermitian, in which case $F(t)\geq 0$ for all $t$. This is also true if we consider the {\it ptp}
propagator $F(r)$ instead, where $r$ is the Euclidian distance on the lattice. 
\newCED{The spin structure defining matrix}{The matrix defining the spin structure}
in the operator $O$ is $\Gamma =\gamma_5,\gamma_{i}$ 
for the $\eta_c$ and $J/\Psi$ states, respectively. The parameter $\lambda$ in the disconnected 
diagram in Fig.~\ref{fig:diagram} stands effectively for the various interactions
that can occur between the two quark loops. Its origins can be a combination of the $U_A(1)$ anomaly effects,
glueball interactions and in the dynamical case -- the propagation of light \newCED{}{hadronic} 
modes. At large distances 
the light modes, if they exist, should dominate in $F(r)$. Since $F(r)$ is nonnegative for all
$r$, it follows that in this case $D(r)$ should also be nonnegative in the large distance limit.
The sign of $D(r=0)$, with the above hermiticity condition on $O$, is strictly negative for
the pseudoscalar (and positive for the vector). It follows that in the dynamical case
$D(r)$ \newCED{will change}{changes} sign for the pseudoscalar,
and indeed we observed this sign flip \cite{mine}.
In the quenched case this sign flip would occur only if there are glueballs lighter than
the charmonium state studied and their signal is stronger than the noise in the data at large distances.    
A simplified form which describes the behavior of $D(r)$ in momentum space is 
\be
D(p^2)\sim \underbrace{\left(C+\frac{f}{p^2+m_l^2}\right)}_\lambda
\left( \frac{a}{p^2+m_c^2} + \frac{b}{p^2+m^{\star 2}_c}\right)^2,
\label{eq:fit_full}
\ee
where we have included in the quark loops one ground state, characterized by mass $m_c$ and an excited one with
mass $m_c^\star$.
The parameter $\lambda$ is represented by a sum of two terms: $C$ stands 
for possible effects of the $U_A(1)$ anomaly (or other effects 
which cannot be decomposed spectrally) and
$f/(p^2+m_l^2)$ is an effective light mode term. \newCED{Fits to Eq.~(\ref{eq:fit_full}) after 
it is Fourier transformed back to space-time on a lattice,
should have the rotational symmetry violations taken into account.}
{If in Eq.~(\ref{eq:fit_full}) we use $4\sin^2(p_\mu a/2)$ in place of $(p_\mu a)^2$, its
discrete Fourier transform accounts for violations of rotational symmetry.
}
As in \cite{mine} the masses
$m_c$ and $m_c^\star$ have been determined from fits to  the connected charmonium propagators 
and are kept fixed.
Then the mass shift due to the disconnected diagram contribution can be approximated as follows
\be
\Delta m =m_c-m_f= \frac{\lambda(-m_c^2)a^2}{\sqrt{32}A_tm_c^2}, 
\ee
where $m_f$ is the full propagator mass and the amplitude $A_t$ is determined 
from the timeslice-to-timeslice connected diagram charmonium propagator. The sign of 
the mass shift depends on the sign of $\lambda(-m_c^2)$.
\section{Dynamical case}
\begin{figure}[ht]
\epsfxsize=95mm
\begin{center}
\epsfbox{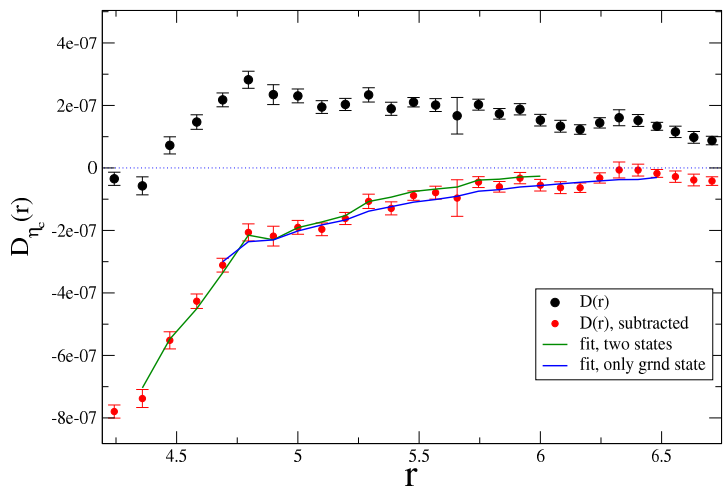}
\caption{The disconnected {\it ptp} propagator for $\eta_c$ in the dynamical case before and 
after the subtraction of the light modes. Also shown are fits to the data after subtraction
with one and two states.}
\label{fig:pssub}
\end{center}
\end{figure}

\begin{figure}[ht]
\epsfxsize=95mm
\begin{center}
\epsfbox{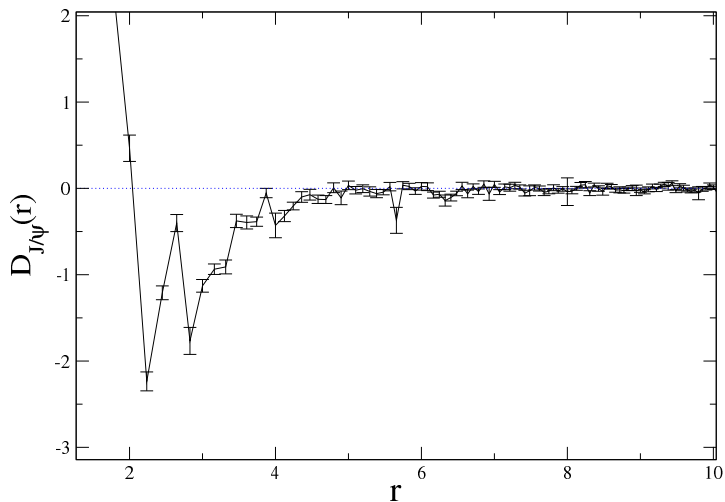}
\caption{The disconnected {\it ptp}  propagator for $J/\Psi$ in the dynamical case.}
\label{fig:vecdyn}
\end{center}
\end{figure}

Fitting our results for $D(r)$ in the dynamical case (for details see \cite{mine}) 
directly to Eq.~(\ref{eq:fit_full})
does not work -- our model most probably has too many parameters and requires 
higher quality statistics. We are forced to 
make further simplifications in our fitting form by removing the light modes from the
data. We subtracted the asymptotic form $Le^{-m_lr}/r^\frac{3}{2}$ with 
values of the parameters $L$ and $m_l$ which we already determined in \cite{mine}.
The data for $D_{\eta_c}(r)$ before and after the subtraction is shown in Fig.~\ref{fig:pssub}.
We fit the subtracted data to the form below, where in this case the absolute value
of $C$ is absorbed in the parameters $a$ and $b$ and only its sign remains:
\be
D(p^2)\sim sign(C)\left( \frac{a}{p^2+m_c^2} + \frac{b}{p^2+m^{\star 2}_c}\right)^2.
\label{eq:fit_s}
\ee
The masses $m_c$ and $m_c^\star$ are the same values used in the fits in \cite{mine}.
Using the above form allows us to obtain a good $\chi^2/{\rm DoF}$, but the fits do 
not deliver consistent results when we change the fitting range or consider
just one ground state. Fig.~\ref{fig:pssub}, where we show two
fits with $\chi^2/{\rm DoF}\approx 1$, illustrates the problem . The first fit (green) 
is a two state fit as in
Eq.~(\ref{eq:fit_s}) and yields $\Delta m_{\eta_c} = -0.7(5)$ MeV. The second fit (blue) is 
a ground state fit only, with the range of $r$ shifted to larger values. This fit
gives $\Delta m_{\eta_c} = -5.5(4)$ MeV. These two values, although
consistent with our rough estimates from \cite{mine}, are not consistent with each other
within the errors obtained from the \newCED{fits. Probably there 
is a systematic error not taken into account.}{fits, suggesting a substantial systematic error.}
%
%
We also have results for $D_{J/\Psi}(r)$, shown in Fig.~\ref{fig:vecdyn}. The signal
for the vector is much more noisy and falls off into the noise at much shorter distances than 
it does in the pseudoscalar case (due to the larger mass of the $J/\Psi$).
We can make only rough estimations for the values of the disconnected diagram in this case:
$-1\,\,{\rm MeV}<\Delta m_{J/\Psi}<0\,\,{\rm MeV}$.
These results show that 
the dynamical case requires a better understanding and further study.

\section{Quenched case}
\begin{figure}[ht]
\epsfxsize=95mm
\begin{center}
\epsfbox{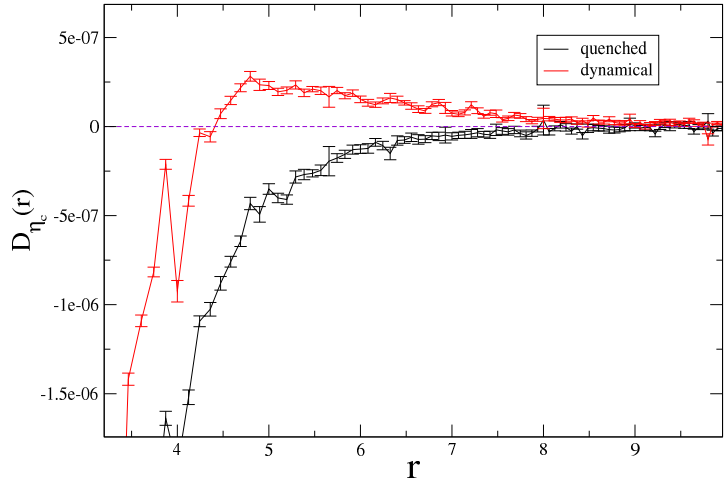}
\caption{Comparison between the $\eta_c$ dynamical and quenched 
disconnected propagators at the same lattice spacing of $a\approx 0.09$ fm.}
\label{fig:pscomp}
\end{center}
\end{figure}

\begin{figure}[ht]
\epsfxsize=95mm
\begin{center}
\epsfbox{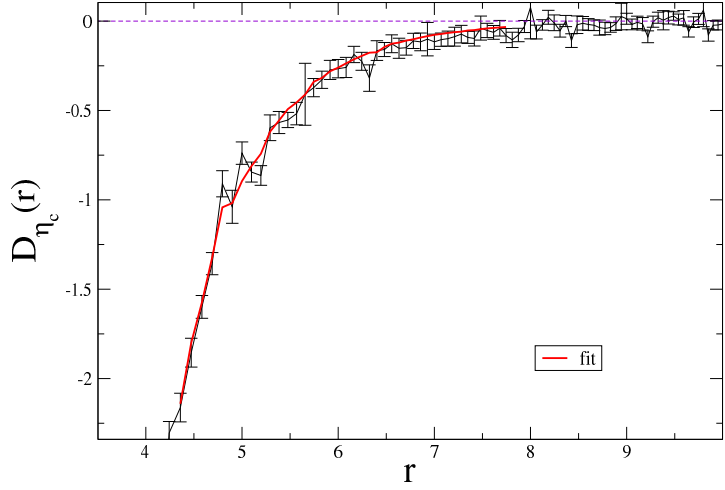}
\caption{The disconnected {\it ptp} propagator for $\eta_c$ on the fine 
quenched lattices ($a\approx 0.09$ fm). }
\label{fig:psquf}
\end{center}
\end{figure}

In the quenched case the behavior of the disconnected propagator is expected to be simplified
due to the absence of light hadronic modes propagating at large distances. 
As in \cite{mine} we use clover fermions to represent
the charm quarks on the lattice. We have results for two
quenched ensembles, ``fine'' and ``superfine'', with lattice spacing of $a\approx 0.09$ fm
and $0.063$ fm, respectively. The lattice volume for the fine ensemble is $28^3\times96$ and
for the superfine it is $48^3\times144$. The respective sizes of the ensembles are 366 and 
124 configurations, and the charm quark kappas are $\kappa=0.127$ and 0.130.
It is interesting to compare $D_{\eta_c}(r)$ on the fine quenched ensemble with the dynamical
result at the same lattice spacing of $a\approx 0.09$ fm. Figure~\ref{fig:pscomp} shows
that the main difference is that the sign of $D_{\eta_c}(r)$ stays constant 
in the region where we see the dynamical data flip sign. We interpret the behavior 
of the quenched data as evidence that in the region where we have a clear signal, 
not only are there no light meson modes due to the quenching, but also if there are glueballs
lighter than the $\eta_c$ in our case at all, their signal is very weak. This makes the use of the
fitting form of Eq.~(\ref{eq:fit_s}) justifiable in this case.
The fits in the quenched case give consistent results under changes of the fit range and 
number of states. The result for $D_{\eta_c}(r)$ on the fine ensemble is 
shown in Fig.~\ref{fig:psquf}.
The fit to $D_{\eta_c}(r)$ is done with $m_c=0.9781$, $m_c^\star=1.330$, known 
with high accuracy from fits
to the connected propagator.
The fitting range is $r=4.3-7.8$  and the fit has $\chi^2/{\rm DoF}=40/40$.
We obtain $a=109(15)$ and $b=294(41)$ for the fit parameters in Eq.~(\ref{eq:fit_s}).
This yields $\Delta m_{\eta_c}= -3.3(9) {\rm MeV}$.

\begin{figure}[ht]
\epsfxsize=95mm
\begin{center}
\epsfbox{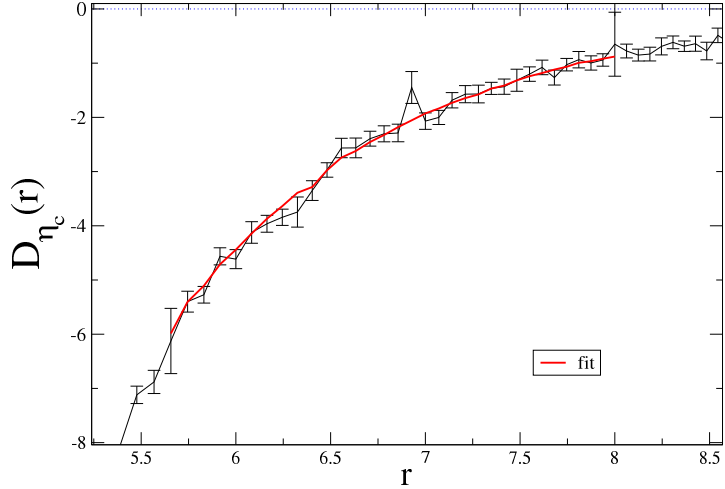}
\caption{The disconnected {\it ptp}  propagator for $\eta_c$ on the superfine 
quenched lattices ($a\approx 0.063$ fm).}
\label{fig:psqusf}
\end{center}
\end{figure}

\begin{figure}[ht]
\epsfxsize=95mm
\begin{center}
\epsfbox{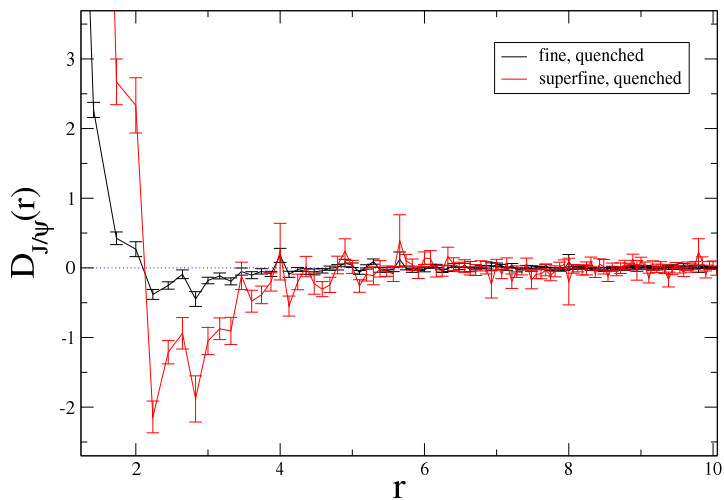}
\caption{The disconnected {\it ptp}  propagator for $J/\Psi$ on the fine and
 superfine quenched lattices.}
\label{fig:vecqu}
\end{center}
\end{figure}

We treat the superfine ensemble results similarly. They are shown in Fig.~\ref{fig:psqusf}.
From the fit which is done with $m_c=0.6509$, $m_c^\star=0.8606$
and fitting range $r=5.6-8$ with $\chi^2/{\rm DoF}=31/31$,
we obtain $a=131(17)$ and $b=246(38)$.
This means $\Delta m_{\eta_c}=-3.1(8) {\rm MeV}$ from the superfine calculation. The results
from both quenched  fine and superfine calculations are consistent with each other 
and with the rough estimates from the dynamical case. They show that
the $\eta_c$ mass is slightly increased due to the disconnected diagram 
contribution. 
We also studied the disconnected propagator for $J/\Psi$ in the quenched case. 
Figure~\ref{fig:vecqu} shows $D_{J/\Psi}(r)$ on both fine and superfine lattices.
The behavior of the data in both cases is similar: the noise is larger than 
it is in the pseudoscalar case, and the signal for the state is overwhelmed
by the noise at shorter $r$. Fits to the data have good $\chi^2/{\rm DoF}$ but the error
on the fit parameters are very large. At this point we can only estimate
that, similarly to the dynamical case, $-1\,\,{\rm MeV}<\Delta m_{J/\Psi}<0\,\,{\rm MeV}$.

\section{Conclusions}
We studied the contributions of the disconnected diagrams to the
masses of $\eta_c$ and $J/\Psi$ in the dynamical and the simplified 
quenched cases. We introduced a new fitting procedure which 
takes into account rotational symmetry violations. It gives consistent results 
with our previous fitting method, but the dynamical case requires further study
to make more accurate predictions. The quenched results 
for $\Delta m_{\eta_c}$ are the same within error
for two lattice spacings $0.09$ and $0.063$ fm: $-3.3(9)$ and $-3.1(8)$ MeV, respectively.
This consistency suggests that the discretization errors are smaller than our statistical errors. 
Our results show that the disconnected diagram contributions increase the $\eta_c$ mass,
which is contrary to the perturbative estimate of a $2.4$ MeV decrease \cite{hisq}.
The mechanism of this increase can be a combination of two effects: mixing 
with glueballs with masses lower than the mass of the $\eta_c$ and the influence 
of the $U_A(1)$ anomaly. In our case the pole mass of the $\eta_c$ is lighter than
the physical one by about $1$ GeV. This is due to the fact that the $\kappa$-tuning 
for the charm quark was done for the kinetic mass instead of the pole mass. 
The lightness of our $\eta_c$ means that there is no lighter 
glueball to mix with \cite{glue} and 
the increase of its mass is due to the anomaly alone. We are currently starting a 
quenched calculation at smaller charm quark kappa, which would give a pole mass
 closer to the physical one for $\eta_c$. We want to \newCED{establish whether}{rule out
the possibility that} the mass
increase in the $\eta_c$ we observe \newCED{is not}{is} an artifact  of the current light
pole mass.
In the case of the vector state $J/\Psi$, in both dynamical and quenched cases,
we can only estimate that its mass will increase by an amount smaller than $1$ MeV 
as a result of the disconnected diagram contributions. We need higher statistics 
to be able to achieve a better estimate since the signal in the vector case 
is noisier and falls off rather quickly with $r$ due to the fact that the vector state is heavier
than the pseudoscalar.
We conclude that as a whole, the hyperfine splitting will decrease by an amount up to a few MeV
when the disconnected diagrams are taken into account in the lattice charmonium 
calculations.
This means that the discrepancies between the lattice calculations (based
solely on connected diagrams) and the experiment are more likely attributable to 
\newCED{the discretization
errors in the actions which represent the heavy quarks on the lattice}
{heavy quark discretization errors}.

\end{document}